\documentclass[prd,twocolumn,showpacs,amsmath,amssymb]{revtex4}

\usepackage{amsfonts}
\usepackage{amssymb}
\usepackage{mathrsfs}
\usepackage{txfonts}
\usepackage{graphicx}
\usepackage{dcolumn}
\usepackage{bm}

\begin{document}

\title{Inflationary spectra from a near $\Omega$-deformed space-time transition point in Loop Quantum Cosmology}

\author{Long Chen}
\email{chen_long@mail.bnu.edu.cn}
\affiliation{Department of Physics, Beijing Normal University, Beijing 100875, China}
\author{Jian-Yang Zhu}
\thanks{Corresponding author}
\email{zhujy@bnu.edu.cn}
\affiliation{Department of Physics, Beijing Normal University, Beijing 100875, China}
\date{\today}
\begin{abstract}
Anomaly-free perturbations of loop quantum cosmology with holonomy corrections reveal an $\Omega$ -deformed spacetime structure,
$\Omega:=1-2\rho/\rho_c$, where $\Omega<0$ indicates a Euclidean-like space and $\Omega>0$ indicates a Lorentz-like space.
It would be reasonable to give the initial value at the spacetime transition point, $\rho=\rho_c/2$,
but we find that it is impossible to define a Minkowski-like vacuum even for large $k$ modes at that time.
However, if we loosen the condition and give the initial value slightly after $\Omega=0$, e.g., $\Omega\simeq 0.2$,
the vacuum state can be well defined and, furthermore the slow roll approximation also works well in that region.
Both scalar and tensor spectra are considered in the framework of loop quantum cosmology with holonomy corrections.
We find that if the energy density is not too small in relation to $\rho_c/2$ when the considered $k$ mode crossing the horizon,
effective theory can give a much smaller scalar power spectrum than classical theory and the spectrum of tensor perturbations could blueshift.
But when compared to other observations, since the energy densities when the modes crossed the horizon were significantly smaller than $\rho_c$,
the results we get agree with previous work in the literature and with the classical inflation theory.
\end{abstract}
\pacs{04.60.Pp, 98.80.Cq, 04.60.Kz}
\maketitle

\section{Introduction}
Loop quantum gravity (LQG) \cite{Han2007,Ashtekar2004,Rovelli2004,Thiemann2007} is a promising quantum theory of gravity which is background independent.
The basic variables of LQG are the holonomies of $SU(2)$ connections and the fluxes of densitized triads.
LQG predicts a quantum geometry whose area, volume, and length operators have discrete eigenvalues.
The dynamics of the theory is not yet complete for the complexity of the Hamiltonian constraints.
Loop quantum cosmology (LQC) \cite{Ashtekar2006,Ashtekar2011,Bojowald2008,Vandersloot}, as the symmetry reduced model of LQG,
uses the quantization method of loop quantum gravity and its key observation of quantum geometry to quantize the cosmological model.
In LQC, the big bang predicted by classical theory is replaced by a big bounce for the existence of a nonezero minimal area spectrum in loop quantum gravity.

To obtain more predictions of LQC and to test the validity of the method of quantization,
the theory of perturbations in the framework of LQC also has been developed.
There exist two kinds of perturbation theories to address this problem in the literature.
The first is the addressed metric method \cite{Ashtekar2013},
while the second one is the anomaly freedom perturbation method in the effective theory of LQC \cite{Bojowald2008b,Cailleteau2012,Cailleteau2012b}.
In this paper, we will consider the second method, developed by Bojowald and his collaborators \cite{Bojowald2008b}.
In the effective theory of LQC, there exist two main quantum corrections to the classical theory:
inverse volume corrections and holonomy corrections.
Since holonomy corrections are more simple and display impressive results at high energy density,
in this paper we focus on the perturbation framework under holonomy corrections developed in \cite{Cailleteau2012,Cailleteau2012b}.
The anomaly-free perturbations of loop quantum cosmology with holonomy corrections reveal an $\Omega$-deformed spacetime structure where
$\Omega:=1-2\rho/\rho_c$ is the coefficient appearing in the expression of the Poisson
bracket between two Hamiltonian constraints \cite{Cailleteau2012}, and $\rho_c$ is the energy
density at the bouncing point which is also the maximum density the Universe could obtain.
When $\rho_c/2<\rho\leqslant\rho_c$, we have $\Omega<0$,
so the Universe's structure is like the Euclidean space (where $\Omega=-1<0$).
Also, $\Omega>0$ when $\rho<\rho_c/2$ would give a spacetime structure like the Lorentzian space (where $\Omega=1>0$).
The signature changes at the transition point $\Omega=0$, which is also called the silent surface.
Because the constraint algebra is more like Euclidean space in the region $\Omega<0$,
the time variable behaves like space variables and, because the dynamical equations of perturbations become elliptic,
the solutions of perturbation equations would face a problem of instability for initial value problems \cite{Barrau2015},
which seems puzzling.
A mixed-type characteristic problem was considered in \cite{Bojowald2015}, but there exist several global problems.
In this paper we want to consider the initial value problem at the transition point $\Omega=0$
in the expanding Universe and discard the evolution before that point.
Note that given initial values in the region where $\Omega<0$ \cite{Mielczarek2014a,Yue2013}
and in the contracting phase\cite{Schander2015} are also considered in the literature.

Inflation, as a necessary supplement to the standard cosmological model,
can solve many long-standing problems such as the horizon problem, the flatness problem, etc.
Using the perturbation theory of cosmology and given a natural initial state, Bunch-Davies vacuum, inflation models can also provide a
natural explanation of the structure formation.
Loop quantum cosmology modifies the evolution of cosmology in classical theory;
thus, which predictions the inflation theory in the framework of LQC could give would be an interesting question.
The power spectra of inflation in LQC with inverse volume corrections were calculated in \cite{Bojowald2011} and were shown to be consistent with the observations. The calculations of the power spectra by the corrections of holonomy were also considered in \cite{Mielczarek2014b,Zhu2015},
where the initial values are given at $\Omega\sim1$ which is far from the transition point $\Omega=0$.
The reason for choosing this initial state is as follows.
In the framework of classical theory, by comparing the observations with the inflation theory,
we find the energy density when crossing the horizon to be quite small compared to the Planck energy $\rho_P$.
Then, since the critical density $\rho_c$ in LQC is generally supposed to be of the order of Planck energy,
one would like to think that the quantum corrections of LQC are not large.
But to make the calculations more precise and self-consistent in the framework of LQC,
it is still preferable to consider the initial values at $\Omega =0$,
and to see whether the quantum corrections at high energy could be omitted.
However by some direct calculations of $z''/z$, where $z=z_S$ or $z_T$ appears in the equations of perturbations,
we will find that
it is impossible to define a well-defined Minkowski (Bunch-Davies) vacuum at the transition point which is bad.
However, if we loosen the initial condition and give it near after $\Omega=0$, e.g. $\Omega\sim 0.2$,
one can define a well-defined vacuum state and the slow roll approximation also works.
For these reasons, we will choose the initial value to be slightly after the transition point in this work.

This paper is organized as follows. In Sec.II, we first review the background equations of motion of LQC with holonomy corrections,
then introduce the slow roll approximations in the epoch of inflation.
In Secs.III and IV, we consider the dynamics of scalar perturbations and tensor perturbations, respectively.
After solving the perturbation equations, we get the power spectra of scalar perturbations and tensor perturbations in Sec.V.
In Sec.VI, we compare the results with classical theory and the observations.
Some conclusions are given in Sec.VII.

\section{\label{sec:level2} Background dynamics and slow roll approximations}
Holonomy corrections in LQC modify Friedmann equations into the following effective equations:
\begin{equation}
H^{2}=\frac{\kappa }{3}\rho (1-\rho /\rho _{c}),  \label{FriedmannEqns1}
\end{equation}%
\begin{equation}
\dot{H}=-\frac{\kappa }{2}\dot{\varphi}^{2}\Omega ,  \label{FriedmannEqns2}
\end{equation}%
\begin{equation}
\ddot{\varphi}+3H\dot{\varphi}+V_{\varphi }=0,  \label{FriedmannEqns3}
\end{equation}%
where $\rho_c$ will be considered to be of the order of $\rho_P$, in this paper, $\rho=\dot\varphi^2/2+V(\varphi)$ is the energy
density of inflaton $\varphi$ and the subscript $\varphi$ in $V_\varphi$ represents the derivative of $V$ by $\varphi$, two
or more $\varphi$s indicates higher derivative.

In the epoch of inflation, the density is dominated by the potential $V(\varphi)$: $\dot\varphi^2/V\ll 1$,
and the first equation (\ref{FriedmannEqns1})then becomes $H^2=\frac{\kappa}{3}V(1-V/\rho_c)$.
By differentiating this equation with respect to time and using Eq.(\ref{FriedmannEqns2}),
one can find $3H\dot\varphi+V'=0$, which means that $\ddot\varphi\ll3H\dot\varphi$ when we compare with Eq.(\ref{FriedmannEqns3}).
Thus, we get the slow roll equations of LQC:
\begin{equation}
H^{2}=\frac{\kappa }{3}V(1-V/\rho _{c}),  \label{SlowRollEqns1}
\end{equation}%
\begin{equation}
\dot{H}=-\frac{\kappa }{2}\dot{\varphi}^{2}\Omega ,  \label{SlowRollEqns2}
\end{equation}%
\begin{equation}
\dot{\varphi}=-\frac{V_{\varphi }}{3H},  \label{SlowRollEqns3}
\end{equation}%
where $\Omega=1-2V/\rho_c=1-2\delta_H$, $\delta_H:=V/\rho_c\leqslant \frac{1}{2}$, and
the slow roll conditions are
\begin{equation}
\epsilon :=\frac{3\dot{\varphi}^{2}}{2V}\simeq \frac{1}{2\kappa }\left(
\frac{V_{\varphi }}{V}\right) ^{2}\frac{1}{1-\delta _{H}}\ll 1,
\label{SlowRollConditions1}
\end{equation}%
\begin{equation}
\delta :=\frac{\ddot{\varphi}}{H\dot{\varphi}}\simeq \frac{1}{2\kappa }\left[
\left( \frac{V_{\varphi }}{V}\right) ^{2}\frac{\Omega }{(1-\delta _{H})^{2}}%
-2\frac{V_{\varphi \varphi }}{V}\frac{1}{1-\delta _{H}}\right] \ll 1.
\label{SlowRollConditions2}
\end{equation}%
The hubble slow roll parameter $\epsilon_H:=-\frac{\dot H}{H^2}\simeq\epsilon\frac{\Omega}{1-\delta_H}$
is smaller than the potential's slow roll parameter $\epsilon$, and the parameter $\eta$ is defined as
$\eta:=\frac{1}{\kappa}\frac{V_{\varphi\varphi}}{V}\frac{1}{1-\delta_H}=\epsilon_H-\delta$.

When considering the dynamics of the perturbations, we will need two variables:
$z_S:=a\dot\varphi/H$ related to the scalar perturbations
and $z_T:=a/\sqrt\Omega$ related to the tensor perturbations,
and the expressions of $z_S''/z_S$ and $z_T''/z_T$ in terms of the conformal time $\tau$,
where $'\equiv \frac{d}{d\tau}=a\frac{d}{d t}$.
Generally, we also hope that some extra requirements \cite{Weinberg} enable us to get simple expressions for these two quantities:
the slow roll parameters must evolve slowly and more precisely, we hope that $\epsilon_H$, $\delta$,
and $\frac{1-\Omega}{\Omega}\epsilon$ evolve much more  slowly than $aH$ (see Eq.(\ref{zspprime}) later).
Compare the evolution of $\epsilon_H$ with $aH$:
\begin{eqnarray}
\frac{\dot{\epsilon}_{H}/\epsilon _{H}}{\dot{(aH)}/(aH)} &=&\frac{2\kappa V}{%
3H^{2}}\left( 1-\delta _{H}\right) \left( \delta +\epsilon _{H}+\frac{%
2\delta _{H}}{\Omega }\epsilon \right)   \nonumber \\
&=&2\left( \epsilon _{H}+\delta +\frac{1-\Omega }{\Omega }\epsilon \right)
\end{eqnarray}%
from which one can find that the value will be infinite if $\Omega\rightarrow 0$, which means that the evolution $\epsilon_H$ is not small.
However if we take $\Omega>0.2$, the value could be smaller than $2\delta+9\epsilon$, satisfying the requirement.
The other two quantities give
\begin{eqnarray}
\frac{\dot{\delta}/\delta }{\dot{(aH)}/(aH)} &=&4\epsilon \left( \frac{%
\Omega }{1-\delta _{H}}+\frac{\delta _{H}}{1-\delta _{H}}\frac{\epsilon }{%
\delta }\right)\nonumber\\
&&\pm \frac{\sqrt{2\kappa \epsilon }}{\kappa ^{2}\delta }\frac{V_{\varphi \nonumber
\varphi \varphi }}{V(1-\delta _{H})^{3/2}}\\
&\ll &1
\end{eqnarray}%
\begin{equation}
\frac{\frac{d}{d t}{\ln(\frac{1-\Omega}{\Omega}\epsilon)}}{\dot{(aH)}/(aH)}=2\frac{3\Omega^2-1}{\Omega(1+\Omega)}\epsilon-2\eta\ll1.
\end{equation}
Generally, $\epsilon$ is of the same order as $\delta$,
so the first gives $\frac{\sqrt{2\kappa\epsilon}}{\kappa^2\delta}\frac{V_{\varphi\varphi\varphi}}{V}\ll1$.
The second requirement gives the same requirement from $\epsilon_H$: $\Omega$ should not be too close to 0.
Note that, in classical theory \cite{Weinberg}, these requirements only give $\frac{\sqrt{2\kappa\epsilon}}{\kappa^2\delta}\frac{V_{\varphi\varphi\varphi}}{V}\ll1$,
so the requirement that $\Omega$ is not too close to 0 is a new requirement from LQC.
It should be noted that, if the requirement that $\delta$ evolves slowly compared to $aH$ is not satisfied,
the calculations of power spectra only under slow roll approximation equations. (\ref{SlowRollConditions1}) and (\ref{SlowRollConditions2}) can be seen in Ref. \cite{Stewart}.
In this paper, for simplicity, we assume that slow roll conditions include these extra requirements (as in most papers).

Now we turn to calculating the expressions for $\frac{z_S''}{z_S}$ and $\frac{z''_T}{z_T}$ under slow roll approximations containing the two extra requirements given above.
We follow the calculation method used by Weinberg \cite{Weinberg} and in Ref. \cite{Riotto}.

First let us consider
\begin{equation}
\frac{z_S''}{z_S}=\frac{d}{d\tau}\left(\frac{z_S'}{z_S}\right)+\left(\frac{z_S'}{z_S}\right)^2.
\end{equation}
From the definition of $\epsilon_H$ and $\delta$, one can get
\begin{equation}
\frac{z_S'}{z_S}=aH(1+\delta+\epsilon_H).
\end{equation}
Then
\begin{eqnarray}\label{zspprime}
\frac{d}{d\tau}\left(\frac{z_S'}{z_S}\right)&=& a\frac{d}{dt}\left[aH(1+\delta+\epsilon_H)\right]  \nonumber \\
&\simeq & a(1+\delta+\epsilon_H)\frac{d}{dt}(aH) \nonumber \\
&\simeq & (aH)^2(1+\delta),
\end{eqnarray}
where in the second equality we have used the extra requirement $\epsilon_H$ and $\delta$ evolve smallly enough compared to $aH$.
Thus, we get $\frac{z_S''}{z_S}=(aH)^2(2+2\epsilon_H+3\delta)$.
The expression of $aH\simeq(1+\epsilon_H)/(-\tau)$ can be found in $\frac{d}{dt}(\frac{1}{aH})=-(1-\epsilon_H)$.
Therefore, we get the final expression for $\frac{z_S''}{z_S}$:
\begin{equation} \label{zS''}
\frac{z_S''}{z_S}\simeq\frac{1}{\tau^2}(\nu_S^2-\frac{1}{4}),
\end{equation}
where $\nu_S:=\frac{3}{2}+2\epsilon_H+\delta$.
$z_S$ also can be expressed with $\tau$ from
$\frac{z_S'}{z_S}=aH(1+\delta+\epsilon_H)\simeq\frac{1}{-\tau}(\nu_S-1/2)$:
\begin{equation} \label{zS}
z_S=\tilde z_S\left(\frac{-\tau}{-\tilde\tau}\right)^{-\nu_S+1/2},
\end{equation}
where $\tilde\tau$ is an arbitrary time which will be fixed later.

The calculations of $z_T$ are almost the same and the results are
\begin{equation}
\frac{z_T''}{z_T}=\frac{1}{\tau^2}(\nu_T^2-\frac{1}{4}),~~
z_T=\tilde z_T\left(\frac{-\tau}{-\tilde\tau}\right)^{-\nu_T+1/2},
\end{equation}
where $\nu_T:=\frac{3}{2}+\frac{1-6\delta_H+6\delta_H^2}{\Omega^2}\epsilon_H$ and the extra requirement
that $\frac{1-\Omega}{\Omega}\epsilon$ evolves slowly enough compared to $a H$ was followed.

\section{\label{sec:level3} Scalar perturbations}
The anomaly-free scalar perturbations of cosmology with holonomy corrections in LQC were derived in \cite{Cailleteau2012},
and we will use their results directly.
When calculating the power spectra of perturbations in inflation,
one requires the action of gauge invariant perturbations of fields which, for scalar perturbations, can be derived by the method used in \cite{Maldacena}.
There are five gauge variant variables $(\psi,\phi,B,E,\delta\varphi)$ in the case of the scalar perturbations.
The idea is that, by choosing a special gauge $\delta\varphi=0=E$
and using the density of momentum and Hamiltonian constraints at first order,
one can express all of the other perturbation fields, i.e. $\phi$ and $B$, in one perturbation field, $\psi$
which, in this gauge is just the gauge invariant quantity $\mathcal {R}$.
Thus, the original action to second order of all gauge dependent perturbation fields becomes the action of $\mathcal {R}$.
In LQC, one also can use this method to obtain the action, which is found to be
\begin{eqnarray} \label{action of scalar perturb}
S^{(2)}_S&=& \int d\tau \int d^3x \left(\frac{1}{2}z_S^2\mathcal{R}'^2-\frac{\Omega}{2}z_S^2(\partial\mathcal{R})^2\right)  \nonumber \\
&=&\int d\tau \int d^3x\left(\frac{1}{2}\upsilon_S'^2-\frac{\Omega}{2}(\partial \upsilon_S)^2+\frac{1}{2}\frac{z_S''}{z_S}\upsilon_S^2\right),
\end{eqnarray}
where $\upsilon_S:=z_S\mathcal{R}$.
One can check to see that, by this action,
the equation of motion of $\upsilon_S$ obtained from the action principle is the same as \cite{Cailleteau2012}:
\begin{equation} \label{eqn of vS}
\upsilon_S''-\Omega\Delta \upsilon_S-\frac{z_S''}{z_S}\upsilon_S=0.
\end{equation}

To quantize the fields $\mathcal{R}$ or $\upsilon_S$, define the momentum field $\pi_S:=\frac{\delta L}{\delta \upsilon_S'}=\upsilon_S'$.
Then, from the commutating relation of $\hat\upsilon_S$ and $\hat\pi_S$, one can get
\begin{equation} \label{brackt of vS}
[\hat \upsilon_S(\vec x,\tau),\hat \upsilon_S'(\vec y,\tau)]=i\delta^{(3)}(\vec x,\vec y),
\end{equation}
where we have chosen $\hbar=1$. Expand $\hat\upsilon_S$ as
\begin{eqnarray}
\hat \upsilon_S(\vec x,\tau)&=& \int \frac{d^3k}{(2\pi)^{3/2}}\hat \upsilon_{S\vec k}(\tau)e^{i\vec k \cdot \vec x} \nonumber \\
&=&\int \frac{d^3k}{(2\pi)^{3/2}} [\hat a_S({\vec k})\upsilon_{Sk}(\tau)e^{i\vec k \cdot \vec x}+\hat a^\dag_S(\vec k)\upsilon^*_{Sk}(\tau)e^{-i\vec k \cdot \vec x}] ,
\nonumber \\
\end{eqnarray}
where, in the second equality, we have made $\upsilon_{S\vec k}=\upsilon_{Sk}$ as in the quantum field theory in Minkowski spacetime,
and the self-adjointness of $\hat \upsilon_S$.
Note that $\hat \upsilon_{S\vec k}(\tau)=\hat a_S(\vec k)\upsilon_{Sk}(\tau)+\hat a^\dag_S(-\vec k)\upsilon^*_{Sk}(\tau)$.
If we require that $\hat a_S$ and $\hat a^\dag_S$ are the annihilation and creation operators satisfying
$[\hat a_S(\vec k), \hat a^\dag_S(\vec k')]=\delta^{(3)}(\vec k, \vec k')$,
the commutating relation in Eq.(\ref{brackt of vS}) would give the same Wronskian condition used in classical theory:
\begin{equation} \label{Wronskian condition}
\upsilon_{Sk}\upsilon'^*_{Sk}-\upsilon^*_{Sk}\upsilon'_{Sk}=i.
\end{equation}
The equation of $\upsilon_{Sk}(\tau)$ can be derived from the Eq.(\ref{eqn of vS}) of $\hat \upsilon_S(\vec x,\tau)$
( by acting on the vacuum state $|0\rangle$ ):
\begin{equation} \label{eqn of vk}
\upsilon_{Sk}''+\omega_{Sk}^2(\tau)\upsilon_{Sk}=0,
\end{equation}
where $\omega_{Sk}^2:=\Omega k^2-\frac{z_S''}{z_S}$.

To define the initial state, which is chosen to be the Minkowski vacuum as in the literature,
one needs to give an initial value to the above equation.
We said in the Introduction that the given initial value at the transition point, $\Omega=0$, is a good choice, but we also said we cannot do this.
The reason for this is that, near $\Omega=0$, the quantity $\frac{z_S''}{z_S}$ is found to be positive, which makes $\omega_{Sk}^2\simeq-\frac{z_S''}{z_S}$ negative for all $k$ modes and any value of $\varphi$ shown below.
From Eqs.(\ref{FriedmannEqns1}-\ref{FriedmannEqns3}), one can find, near $\Omega=0$,
\begin{equation}
\frac{z_S''}{z_S}\simeq \kappa a^2\rho_c\left[\frac{1}{6}+12\left(\delta_H-\frac{1}{2}\right)^2-\frac{V_{\varphi\varphi}}{\kappa \rho_c}\right].
\end{equation}
Generally, the term $V_{\varphi\varphi}/\kappa \rho_c$ on the right hand side in the above equation are quite small --
e.g., the model with potential $V=\frac{1}{2}m^2\varphi^2$, where $m\sim10^{-6}m_P$ gives
$V_{\varphi\varphi}/\kappa \rho_c\sim 10^{-12}$ --
so the other two terms always make $z_S''/z_S$  positive.
This would be a bad result, and whether there exists some approach for addressing this problem should be analyzed in future work.
In this paper, we instead consider a loose version of the problem:
the initial value given is near after the transition point.
Fortunately, in the region of low values of $\Omega$, $\omega_{Sk}^2$ can be positive for large $k$ modes.
 Furthermore, we know from Sec.II that, if we take $\Omega\sim0.2$, slow roll approximations also are valid,
thus the general approach of calculating power spectra can be used in LQC.
For these reasons, we will take our initial values at the point $\Omega=0.2$.

As in classical theory, we consider those modes where, initially $\sqrt{\Omega}k\gg aH$.
Since $\sqrt{\Omega}=\sqrt{1-2V/\rho_c}$ varies slowly in the epoch of inflation,
$\omega_{Sk}\simeq\sqrt\Omega k$ also varies slowly and
the solution would be $\upsilon_{Sk}\sim c_1e^{-i\omega_{Sk}(\tau-\tau_*)}+c_2e^{i\omega_{Sk}(\tau-\tau_*)}$,
where $\tau_*$ is the initial time with $\Omega=0.2$.
The positive frequency solution is used to define the Minkowski vacuum,
where the coefficient $c_1$ can be fixed by the Wronskian condition (\ref{Wronskian condition}):
\begin{equation} \label{initial value of vk}
\upsilon_{Sk}\sim\frac{1}{\sqrt{2\sqrt\Omega k}}e^{-i\sqrt\Omega k(\tau-\tau_*)}.
\end{equation}

On the other hand, from the expression in Eq.(\ref{zS''}),
the dynamical equation of $\upsilon_{Sk}$ in Eq.(\ref{eqn of vk}) becomes
\begin{equation} \label{eqn of vk2}
\upsilon_{Sk}''+\left[\Omega {k}^2-\frac{1}{\tau^2}\left(\nu_S^2-\frac{1}{4}\right)\right]\upsilon_{Sk}=0,
\end{equation}
whose general approximated solution (this approximation is appropriate since $\Omega$ varies slowly) is
\begin{equation} \label{general solution of vk}
\upsilon_{Sk}(\tau)=\sqrt{-\tau}\left[d_1(k)H^{(1)}_{\nu_S}\left(-\sqrt\Omega k\tau\right)
+d_2(k)H^{(2)}_{\nu_S}\left(-\sqrt\Omega k\tau\right)\right],
\end{equation}
where $H^{(1)}_\nu$ and $H^{(2)}_\nu$ are the Hankel's functions of first and second class, respectively.
To fix the coefficients $d_1$ and $d_2$, one needs to use the initial state in Eq.(\ref{initial value of vk}), where $\sqrt\Omega k(-\tau)\simeq\frac{\sqrt\Omega k}{aH}\gg1$.
From the asymptotic properties of $H^{(1)}_\nu$ and $H^{(2)}_\nu$,
\begin{equation}
H^{(1)}(x\gg1)\simeq\sqrt{\frac{2}{\pi x}}e^{i(x-\frac{\pi}{2}\nu-\frac{\pi}{4})},
\end{equation}
\begin{equation}
H^{(2)}(x\gg1)\simeq\sqrt{\frac{2}{\pi x}}e^{-i(x-\frac{\pi}{2}\nu-\frac{\pi}{4})},
\end{equation}
and by comparing them to the initial value, one can get
\begin{eqnarray}
d_1=\frac{\sqrt\pi}{2}e^{i\frac{\pi}{2}\left(\nu_S+\frac{1}{2}\right)+i\sqrt\Omega k\tau_*},~~
d_2=0.
\end{eqnarray}
Thus we get the solution for $\upsilon_{Sk}(\tau)$ with the initial value (\ref{initial value of vk}) and obtain the solution for $\mathcal{R}_k$:
\begin{equation}\label{solution of Rk}
\mathcal{R}_k(\tau)=\frac{\sqrt\pi}{2}e^{i\frac{\pi}{2}\left(\nu_S+\frac{1}{2}\right)+i\sqrt\Omega k\tau_*}
\frac{\sqrt{-\tau}}{z_S}H^{(1)}_{\nu_S}\left(-\sqrt\Omega k\tau\right).
\end{equation}

To compare this with observations, we want to know, when $\sqrt \Omega k\ll aH\simeq\frac{1}{-\tau}$, i.e. $x\ll 1$, how $\mathcal{R}_k(\tau)$ will behave.
From the asymptotic property
$H^{(1)}_\nu(x\ll 1)\simeq \sqrt{\frac{2}{\pi}}e^{-i\pi/2}2^{\nu-3/2}\frac{\Gamma(\nu)}{\Gamma(3/2)}x^{-\nu}$,
one can get when $\sqrt\Omega k\ll aH$,
\begin{eqnarray}
\mathcal{R}_k&=& \frac{1}{\sqrt{2\sqrt\Omega k}}\frac{ie^{i\sqrt\Omega k\tau_*}}{z_S}(-\sqrt\Omega k\tau)^{\frac{1}{2}-\nu_S} \nonumber \\
&=&\frac{ie^{i\sqrt\Omega k\tau_*}}{\sqrt2\tilde z_S (\tilde a\tilde H)^{-\nu_S+1/2}}(\sqrt{\Omega(\tau)}k)^{-\nu_S},
\end{eqnarray}
where we have used the expression for $z_S$ in Eq.(\ref{zS}).
Note that because $\Omega$ depends slightly on time at the high energy density stage,
the expression of $\mathcal {R}_k$ is not independent of time.
However, when the Universe evolves into the low density region -for example, when inflation ends, we have $\Omega\simeq 1$ and
the value of $\mathcal{R}_k$ becomes time independent:
$\mathcal{R}_k=\frac{ie^{i k\tau_*}}{\sqrt2\tilde z_S (\tilde a\tilde H)^{-\nu_S+1/2}}k^{-\nu_S}$.
Before fixing the time $\tilde \tau$, we note that
$|\mathcal{R}_k|\propto k^{-\nu_S}$ which will be useful for giving the spectral index when calculating the power spectrum of scalar perturbations later.

At last, we fix the time $\tilde \tau$, as in the literature,
at $\tau_k$ when the $k$ mode crosses the Hubble horizon: $\sqrt{\Omega_k}k=a_kH_k$,
from which one can get
$\tilde z_S=\frac{\tilde a \dot{\tilde \varphi}}{\tilde H}
=\pm a_k\sqrt{\frac{2\epsilon_{Hk}}{\kappa \Omega_k}}
=\pm\frac{k}{H_k}\sqrt{\frac{2\epsilon_{Hk}}{\kappa}}$.
Thus the final expression of $\mathcal {R}_k$ is
\begin{equation}
\mathcal{R}_k=\pm\frac{i\sqrt\kappa e^{i k\tau_*}}{2}\frac{H_k}{\sqrt{\epsilon_{Hk}}}\Omega_k^{\frac{\nu_S-1/2}{2}}k^{-3/2}.
\end{equation}
It may be noted that the information on choosing the initial condition $\Omega_*=0.2$ only enters into the exponent term $e^{ik\tau_*}$;
thus, when considering the power spectrum of scalar perturbations where we only need the absolute value of $\mathcal{R}_k$,
the true value of $\Omega_*$ would not be relevant.
We will use this expression to define the power spectrum of the perturbations which will be shown in Sec. V.

\section{\label{sec:level4} Tensor perturbations}
In this section, we consider tensor perturbations in loop quantum cosmology.
The tensor perturbations or gravitons $h^a_i$ \cite{Bojowald2008c} with the traceless and transverse
conditions $h^i_i=0=\partial_ah^a_i$ are already gauge invariant.
The theory of tensor perturbations in LQC was derived in \cite{Cailleteau2012b}.
The approach for getting the quantum corrected Hamiltonian constraints is to make a consistent framework in all types of perturbations;
that is to say, we should make different types of perturbations share the same coefficients with the counterterms in the expansion of Hamiltonian constraints.
Since the scalar perturbations almost determine the coefficients and the counterterms, we get the desired results for the tensor perturbations.
The action can also be obtained in this framework:
\begin{equation} \label{action of tensor perturb}
S^{(2)}_T=\frac{1}{8\kappa}\int d\tau \int d^3xz^2_T\left[{h^i_a}'{h^a_i}'-\Omega(\partial_dh^i_a)(\partial_dh^a_i)\right].
\end{equation}
From this action, one can get the dynamics of $h^a_i$:
\begin{equation} \label{eqn of h}
{h^a_i}''+\left(2\mathcal{H}-\frac{\Omega'}{\Omega}\right){h^a_i}'-\Omega\Delta h^a_i=0.
\end{equation}

The traceless and transverse conditions make $h^a_i$ only have 2 independent degrees of freedom (see \cite{Weinberg}),
and one can expand the tensor perturbations as (see also \cite{Riotto})
\begin{eqnarray}
\hat h_{ij}(\vec x,\tau)
&=&\sum _{\sigma=\pm2}\int \frac{d^3k}{(2\pi)^{3/2}}e^\sigma_{ij}(\hat k)\hat h^\sigma_{\vec k}(\tau)e^{i\vec k\cdot \vec x} \nonumber \\
&=&\sum _{\sigma=\pm2}\int \frac{d^3k}{(2\pi)^{3/2}}
\left[h^\sigma_k(\tau)e^\sigma_{ij}(\hat k)\hat a^\sigma_T(\vec k)e^{i\vec k \cdot \vec x}\right. \nonumber \\
&&\left.+ h^{\sigma*}_k(\tau)e^{\sigma*}_{ij}(\hat k)\hat a^{\sigma\dag}_T(\vec k)e^{-i\vec k \cdot \vec x}\right],
\end{eqnarray}
where $\hat h^\sigma_{\vec k}(\tau)=h^\sigma_k(\tau)e^\sigma_{ij}(\hat k)\hat a^\sigma_T(\vec k)
+ h^{\sigma*}_k(\tau)e^{\sigma*}_{ij}(\hat k)\hat a^{\sigma\dag}_T(-\vec k)$,
$\hat k=\vec k/|\vec k|$.
The parameter $\sigma=\pm2$ is called the helicity of gravitons, and $e^\sigma_{ij}$ satisfies
\begin{equation} \label{relation of eij}
\begin{array}{c}
e^\sigma_{ij}=e^\sigma_{ji},\;~~e^\sigma_{ii}=0=k^ie^\sigma_{ij},\\
e^\sigma_{ij}(-\hat k)=e^{\sigma*}_{ij}(\hat k),\;~~e^{\sigma*}_{ij}(\hat k)e^{\tilde{\sigma}}_{ij}(\hat k)=2\delta_{\sigma,\tilde\sigma}.
\end{array}
\end{equation}
To get the Wronskian condition of $h^\sigma_k$, one can rewrite $\hat h^\sigma_{\vec k}$ into two self-adjoint operators, $\hat{\mathcal{U}}^\sigma_{\vec k}$ and
 $\hat{\mathcal{V}}_{\vec k}$ (which are real quantities in classical theory):
$\hat h^\sigma_{\vec k}=\hat{\mathcal{U}}^\sigma_{\vec k}+i\hat{\mathcal{V}}_{\vec k}$.
By using the relations of Eq.(\ref{relation of eij}), one can get the action of $\mathcal{U}_{\vec k}$ and $\mathcal{V}_{\vec k}$
from the original action in Eq.(\ref{action of tensor perturb}).
Then their commuting relations, together with $[\hat a^\sigma_T(\vec k), \hat a^{\sigma'\dag}_T(\vec k')]
=\delta_{\sigma,\sigma '}\delta^{(3)}(\vec k, \vec k')$, will give the Wronskian condition of $h^\sigma_k$:
$h^\sigma_k{h^{\sigma*}_k}'-h^{\sigma*}_k{h^\sigma_k}'=\frac{2i\kappa\Omega}{a^2},\sigma=\pm2.$
Defining $\upsilon_{Tk}:=\frac{z_T}{\sqrt{2\kappa}}h^\sigma_k$, $\forall \sigma$, the above relation then becomes
\begin{equation} \label{Wronskian condition2}
\upsilon_{Tk}\upsilon'^*_{Tk}-\upsilon^*_{Tk}\upsilon'_{Tk}=i.
\end{equation}
It is interesting to note that there exists another method for getting this relation showed in \cite{Mielczarek2014b}, where the author used an effective 4-metric.
The equation of $\upsilon_{Tk}$ also can be derived from the equation of $h^a_i$ in Eq.(\ref{eqn of h}):
\begin{equation} \label{eqn of vTk}
\upsilon_{Tk}''+\omega^2_{Tk}\upsilon_{Tk}=0,
\end{equation}
where $\omega^2_{Tk}=\Omega k^2-\frac{z_T''}{z_T}$.
As in the case of the scalar perturbations, we also find that $\frac{z_T''}{z_T}$ is positive near $\Omega\sim0^+$:
\begin{equation*}
\frac{z_T''}{z_T}\simeq \frac{9}{4}\kappa\rho_ca^2\left(\frac{1-\frac{2V}{\rho_c}}{1-\frac{2\rho}{\rho_c}}\right)^2.
\end{equation*}
Worse than the case of scalar perturbations, this quantity would be infinite except when $\dot\varphi=0$.
Thus, we cannot define a Minkowski vacuum at $\Omega=0$-but define it after that time.

Since the calculations below are similar to Sec.III, we only give the main results.
The solution of $h^\sigma_k=\sqrt{2\kappa}\frac{v_{Tk}}{z_T}$ is
\begin{equation}\label{solution of hk}
h^\sigma_k(\tau)=\sqrt{\frac{\pi\kappa}{2}}e^{i\frac{\pi}{2}\left(\nu_T+\frac{1}{2}\right)+i\sqrt\Omega k\tau_*}
\frac{\sqrt{-\tau}}{z_T}H^{(1)}_{\nu_T}\left(-\sqrt\Omega k\tau\right), \forall \sigma.
\end{equation}
When $\sqrt\Omega k\ll aH$, it can be approximated as
\begin{eqnarray} \label{solution of hk2}
h^\sigma_k(\tau)&=& \frac{i\sqrt\kappa e^{ik\tau_*}}{\tilde z_T (\tilde a\tilde H)^{-\nu_T+1/2}}k^{-\nu_T}  \nonumber \\
&=&i\sqrt\kappa e^{ik\tau_*}\frac{H_k\Omega_k^{\frac{\nu_T-1/2}{2}}}{k^{3/2}},
\end{eqnarray}
where, in the first equality, the time $\tau$ is taken at the low energy density region where $\Omega\simeq1$,
and, in the second equality, we choose $\tilde\tau=\tau_k$, such that $\sqrt\Omega_k k=a_kH_k$,
which gives $\tilde z_T=\frac{\tilde a}{\sqrt{\tilde\Omega}}=\frac{k}{H_k}$.
Note that, from the first equality, one can find that $|h^\sigma_k|\propto k^{-\nu_T}$, which will be used to determine the spectral index of the tensor perturbations.

\section{\label{sec:level5}Power spectra of inflations}
With the solutions for the scalar and tensor perturbations obtained in Secs.III and IV,
we can now get the power spectra of these modes.

First, we consider the scalar perturbations.
The power spectrum of $\mathcal{R}$ is defined as $P_{\mathcal{R}}:=\frac{k^3}{2\pi^2}|\mathcal{R}_k(\tau)|^2$ from the two point correlation function:
\begin{eqnarray}
\langle 0|\hat{\mathcal{R}}(\vec x,\tau)\hat{\mathcal{R}}(\vec y,\tau)|0\rangle
&=&\int \frac{d^3k}{(2\pi)^3}|\mathcal{R}_k(\tau)|^2e^{i\vec k \cdot (\vec x-\vec y)} \nonumber \\
&=&\int\frac{k^2\sin\theta dkd\theta d\varphi}{(2\pi)^3}|\mathcal{R}_k(\tau)|^2e^{ikr\cos\theta} \nonumber \\
&=&\int\frac{dk}{k}\frac{\sin(kr)}{kr}\frac{k^3}{2\pi^2}|\mathcal{R}_k(\tau)|^2 \nonumber \\
&=&:\int\frac{dk}{k}\frac{\sin(kr)}{kr}P_{\mathcal{R}}(k,\tau).
\end{eqnarray}
From the result of Sec. III, we have arrived at the expression for $\mathcal{R}_k(\tau)$ when $\sqrt\Omega k\ll aH$ and $\Omega\simeq1$,
so the power spectrum is found to be
\begin{equation}
P_{\mathcal{R}}(k)=\frac{\kappa}{8\pi^2}\frac{H_k^2}{\epsilon_{Hk}}\Omega_k^{\nu_S-\frac{1}{2}}\propto k^{3-2\nu_S},
\end{equation}
which is time independent and almost scale invariant.
Thus, we get the power spectrum
\begin{eqnarray}\label{spectrum of scalar perturb}
A_S&=& \frac{\kappa}{8\pi^2}\frac{H^2}{\epsilon_H}\Omega^{\nu_S-\frac{1}{2}} \nonumber \\
&=&\frac{\kappa}{8\pi^2}\frac{H_\mathfrak{c}^2}{\epsilon_\mathfrak{c}}(1-\delta_H)^3(1-2\delta_H)^{3\epsilon_H-\eta} \nonumber \\
&\simeq&\frac{\kappa}{8\pi^2}\frac{H_\mathfrak{c}^2}{\epsilon_\mathfrak{c}}(1-\delta_H)^3,
\end{eqnarray}
where, in the last equality, $(1-2\delta_H)^{3\epsilon_H-\eta}\simeq1$ was used for the smallness of $\epsilon_H,\eta$,
and we have defined the classical expressions denoted with subscript $\mathfrak{c}$ as
\begin{equation} \label{classical expr}
\epsilon_\mathfrak{c}:=\frac{1}{2\kappa}\left(\frac{V'}{V}\right)^2,~~
\eta_\mathfrak{c}:=\frac{1}{\kappa}\frac{V''}{V},~~
H^2_\mathfrak{c}:=\frac{\kappa}{3}V.
\end{equation}
The spectral index $n_S$ is found to be
\begin{eqnarray} \label{index of scalar perturb}
n_S-1&=& 3-2\nu_S \nonumber \\
&=&-6\epsilon_\mathfrak{c}\frac{1-2\delta_H}{(1-\delta_H)^2}+2\eta_\mathfrak{c}\frac{1}{1-\delta_H}.
\end{eqnarray}

The calculations of the tensor perturbations are similar to the scalar case.
First the power spectrum
$P_T:=\frac{2k^3}{\pi^2}|h^\sigma_k|^2$ is defined from the correlation functions
\begin{eqnarray}
\langle 0|\hat h_{ij}(\vec x,\tau)\hat h_{ij}(\vec y,\tau)|0\rangle
&=& \underset{\sigma}\sum \int \frac{d^3k}{(2\pi)^3}|h^\sigma_k|^2e^\sigma_{ij}(\hat k)e^{\sigma*}_{ij}(\hat k)e^{i\vec k \cdot (\vec x-\vec y)}\nonumber \\
&=&4\int \frac{d^3k}{(2\pi)^3}|h^\sigma_k(\tau)|^2e^{i\vec k \cdot (\vec x-\vec y)}\nonumber \\
&=&\int\frac{dk}{k}\frac{\sin(kr)}{kr}\frac{2k^3}{\pi^2}|h^\sigma_k(\tau)|^2 \nonumber \\
&=&:\int\frac{dk}{k}\frac{\sin(kr)}{kr}P_T(k,\tau),
\end{eqnarray}
where the property of $e^\sigma_{ij}$ in Eq.(\ref{relation of eij}) was used to get the second equality.
By using the solution Eq.(\ref{solution of hk2}), one can find when $k\ll aH$,
\begin{equation}
P_T(k)=\frac{2\kappa}{\pi^2}H_k^2\Omega_k^{\nu_T-\frac{1}{2}}\propto k^{3-2\nu_T}.
\end{equation}
Then we get the power spectrum of the tensor,
\begin{equation} \label{spectrum of tensor pertub}
A_T=\frac{2\kappa}{\pi^2}H^2\Omega^{\nu_T-\frac{1}{2}}
=\frac{2\kappa}{\pi^2}H_{\mathfrak{c}}^2(1-\delta_H)(1-2\delta_H)^{\nu_T-\frac{1}{2}},
\end{equation}
and its spectral index,
\begin{eqnarray}\label{index of tensor pertub}
n_T=3-2\nu_T=-2\epsilon_{\mathfrak{c}}\frac{1-6\delta_H+6\delta^2_H}{(1-2\delta_H)(1-\delta_H)^2}.
\end{eqnarray}
The tensor-scalar ratio of power spectra is
\begin{equation} \label{ratio of t-s}
r:=\frac{A_T}{A_S}=16\epsilon_H=16\epsilon_{\mathfrak{c}}\frac{1-2\delta_H}{(1-\delta_H)^2}.
\end{equation}

\section{\label{sec:level6}Comparison with classical theory and observations}
In this section, we first compare the results given in the previous section with the predictions of classical theory,
then compare them with observations.

Classical theory predicts the power spectra of perturbations and their spectral indexes as follows \cite{Weinberg,Riotto}
\begin{equation} \label{}
A^\mathfrak{c}_S =\frac{\kappa}{8\pi^2}\frac{H_\mathfrak{c}^2}{\epsilon_\mathfrak{c}},~~r^\mathfrak{c}=16\epsilon_{\mathfrak{c}},
\end{equation}
and
\begin{equation} \label{}
n^\mathfrak{c}_S-1=-6\epsilon_\mathfrak{c}+2\eta_\mathfrak{c},~~n^\mathfrak{c}_T=-2\epsilon_{\mathfrak{c}}.
\end{equation}
When making a comparison with classical theory, for explicitness, we suppose that the values of the inflation field when crossing the Hubble horizon are the same for quantum and classical theory.
Then,
\begin{enumerate}
  \item The power spectrum $A_S$ of scalar perturbations in LQC in Eq.(\ref{spectrum of scalar perturb}) is smaller than in classical theory for the quantum correction term
  $(1-\delta_H)^3<1$, where $\delta_H\in(0,0.4)$.
  If $\frac{V_{exit}}{\rho_c}\simeq0.4$, the correction of LQC could be large: $A_S\simeq0.2A^{\mathfrak{c}}_S$.
  \item The spectral index $n_S$ of scalar perturbations of LQC is larger than in classical theory,
  which means LQC gives a flater spectrum of scalar perturbations.
  Since the quantum corrections in Eq.(\ref{index of scalar perturb}) are
  $\frac{5}{9}<\frac{1-2\delta_H}{(1-\delta_H)^2}<1$ and $\frac{5}{3}>\frac{1}{1-\delta_H}>1$, we have
  $1-n_S\in\left(\frac{10}{3}(\epsilon_\mathfrak{c}-\eta_\mathfrak{c}),6\epsilon_\mathfrak{c}-2\eta_\mathfrak{c}\right)$.
  \item An interesting thing is that the spectrum of tensor perturbation could be
  either blueshifted if the $k$ mode crosses the horizon at high energy density
  or redshifted if it crosses the horizon at low energy density
  which can be seen from the term in $n_T$ Eq.(\ref{index of tensor pertub}):
  $1-6\delta_H+6\delta^2_H\in(-0.44,1)$.
  \item The tensor-scalar ratio of spectra $r$ is generally smaller than classical theory since the correction term in Eq.(\ref{ratio of t-s}) $\frac{5}{9}<\frac{1-2\delta_H}{(1-\delta_H)^2}<1$.
\end{enumerate}

We have found that, as in classical theory, the power spectra only depends on the values of the quantities when the modes cross the Hubble horizon,
and the information of the initial state is not directly relevant.
Thus, if the quantum corrections of LQC when the $k$ mode crossing the Hubble horizon is small,
the quantum correction to power spectra would be small.
We find that the modes connected with observations \cite{Planck} are in a region with low energy density,
so holonomy corrections to power spectra are quite small, and the results are the same with the results
obtained by Mielczarek \cite{Mielczarek2014b} and in classical theory.
This can be seen as below.
From the observation of the power spectrum of scalar perturbations $A_S\sim10^{-9}$, and in comparison to Eq.(\ref{spectrum of scalar perturb}),
one can find that $\frac{V}{\rho_P}=\frac{3\epsilon_\mathfrak{c}}{8(1-\delta_H)^3}A_S<2\times10^{-10}$,
where we have used the largest possible value of slow roll parameter $\epsilon_\mathfrak{c}=0.1$ and take $\delta_H=0.4$.
If $\rho_c$ is of the same order of $\rho_P$, which is often assumed in the literature,
then we know the energy density is quite low compared to $\rho_c$.

\section{\label{sec:level7} conclusions}
In this paper, we considered the power spectra of scalar perturbations and tensor perturbations in the framework of loop quantum cosmology with holonomy corrections.

Since the dynamical equations of both types of perturbations turn elliptic in the region $\rho>\rho_c/2$,
given that the initial values at those regions would present the problem of instability.
We tried to consider the initial value problem at $\Omega=0$ in the expanding phase,
but found that one cannot define the Minkowski vacuum as in classical theory.
Thus, we instead considered a loose version of this problem, the problem with the initial value slightly after the transition point.
The vacuum state can be well defined for large $k$ modes and the slow roll approximations also can be used which enable us to solve these problems analytically.
Actions of perturbations were obtained which give Wronskian conditions for defining the quantum theory of perturbations.
We found that holonomy corrections can modify the power spectra of the perturbations and their indexes largely when the density is of the order of $\rho_c$.
For example, the power spectrum of scalar perturbations in loop quantum cosmology can be smaller than classical theory in one order,
and the spectrum of tensor perturbations even could blueshift.
When making a connection with observations, however, it was found that the modes we observed are quite low compared to $\rho_c$, which means that quantum corrections from
loop quantum cosmology are small. The results obtained in this paper thus would be the same as the calculations in \cite{Mielczarek2014b}
and the results of classical theory.

In this paper, we found that one cannot define the vacuum state at $\Omega=0$ in the framework of holonomy corrections
since $\frac{z''}{z}$'s are always positive at that point.
Whether things could change if we consider holonomy corrections together with inverse volume corrections is an interesting question.
This perturbation theory of considering these two corrections was derived in \cite{Cailleteau2014},
so the calculations may be possible and deserve to research.

\acknowledgments This work was supported by the National Natural Science Foundation of China (Grants No. 11175019, No. 11575270, and No. 11235003).

\end{document}